\documentclass[10pt,aps,prl,preprint,superscriptaddress]{revtex4-1}
\usepackage{graphicx}
\usepackage{color}
\usepackage{amsfonts}

\begin{document}

\title{Epitaxial Growth of Single-Orientation High-Quality MoS$_2$ Monolayers}

\author{Harsh Bana}
\author{Elisabetta Travaglia}
\affiliation{Department of Physics, University of Trieste, Via Valerio 2, 34127 Trieste, Italy.}

\author{Luca Bignardi}
\author{Paolo Lacovig}
\affiliation{Elettra - Sincrotrone Trieste S.C.p.A., AREA Science Park, Strada Statale 14, km 163.5, 34149 Trieste, Italy.}

\author{Charlotte E. Sanders}
\author{Maciej Dendzik}
\author{Matteo Michiardi}
\author{Marco Bianchi}
\affiliation{Department of Physics and Astronomy, Interdisciplinary Nanoscience Center (iNANO), Aarhus University, Ny Munkegade 120, 8000 Aarhus C, Denmark.}
\author{Daniel Lizzit}
\affiliation{Elettra - Sincrotrone Trieste S.C.p.A., AREA Science Park, Strada Statale 14, km 163.5, 34149 Trieste, Italy.}

\author{Francesco Presel}
\affiliation{Department of Physics, University of Trieste, Via Valerio 2, 34127 Trieste, Italy.}

\author{Dario De Angelis}
\affiliation{Department of Physics, University of Trieste, Via Valerio 2, 34127 Trieste, Italy.}

\author{Nicoleta Apostol}
\affiliation{National Institute of Materials Physics, Atomistilor Str. 405A, 077125, Magurele, Romania.}
\author{Pranab Kumar Das}
\affiliation{Abdus Salam International Centre for Theoretical Physics, Str. Costiera 11, 34151 Trieste, Italy.}
\affiliation{IOM-CNR, Laboratorio TASC, AREA Science Park, Strada Statale 14, km 163.5, 34149 Trieste, Italy.}
\author{Jun Fujii}
\author{Ivana Vobornik}
\affiliation{IOM-CNR, Laboratorio TASC, AREA Science Park, Strada Statale 14, km 163.5, 34149 Trieste, Italy.}

\author{Rosanna Larciprete}
\affiliation{CNR-Institute for Complex Systems, Via dei Taurini 19, 00185 Roma, Italy.}

\author{Alessandro Baraldi}
\affiliation{Department of Physics, University of Trieste, Via Valerio 2, 34127 Trieste, Italy.}
\affiliation{Elettra - Sincrotrone Trieste S.C.p.A., AREA Science Park, Strada Statale 14, km 163.5, 34149 Trieste, Italy.}
\affiliation{IOM-CNR, Laboratorio TASC, AREA Science Park, Strada Statale 14, km 163.5, 34149 Trieste, Italy.}

\author{Philip Hofmann}
\email{correspondence to: philip@phys.au.dk}
\affiliation{Department of Physics and Astronomy, Interdisciplinary Nanoscience Center (iNANO), Aarhus University, Ny Munkegade 120, 8000 Aarhus C, Denmark.}

\author{Silvano Lizzit}
\email{correspondence to: lizzit@elettra.eu}
\affiliation{Elettra - Sincrotrone Trieste S.C.p.A., AREA Science Park, Strada Statale 14, km 163.5, 34149 Trieste, Italy.}

\begin{abstract}
We present a study on the growth and characterization of high-quality single-layer MoS$_2$ with a single orientation, i.e. without the presence of mirror domains. This single orientation of the MoS$_2$ layer is established by means of x-ray photoelectron diffraction. The high quality is evidenced by combining scanning tunneling microscopy with x-ray photoelectron spectroscopy measurements. Spin- and angle-resolved photoemission experiments performed on the sample revealed complete spin-polarization of the valence band states near the K and -K points of the Brillouin zone.  These findings open up the possibility to exploit the spin and valley degrees of freedom for  encoding and processing information in devices that are based on epitaxially grown materials.
\end{abstract}


\maketitle
\section{Introduction}

Novel two-dimensional materials form the basis of future electronic devices that exploit the valley \cite{Mak:2012aa,Zeng:2012aa,Cao:2012ab,Mak:2014aa,Wang2017} and spin \cite{Xiao:2012ab,Xu:2014ac} degrees  of freedom. Single-layer (SL) transition metal dichalchogenides (TMDCs) are particularly promising for such applications because, unlike graphene, their structure breaks inversion symmetry and integrates atoms with a strong spin-orbit interaction. However, exploiting these new degrees of freedom in an actual electronic device requires a distinction between the $K$ and $-K$ points of the Brillouin zone and thus a single orientation of the layer.
Current methods of chemical vapour deposition have not been able to achieve this and have produced mirror twin domains \cite{Zande:2013aa,Zhou:2013ad,Najmaei_2013}. Here we report a protocol for the synthesis of SL MoS$_2$ of a \emph{single} domain orientation. We demonstrate the structural properties of the MoS$_2$ layer using photoelectron diffraction and we measure the complete spin polarization of the valence band states near $K$ and $-K$ by spin- and angle-resolved photoemission spectroscopy.

Early successes in fabricating electronic devices based on single-layer transition metal dichalcogenides took advantage of the direct band gap in SL MoS$_2$ \cite{Mak2010} and WS$_2$, which guarantees large on-off current ratios in field effect transistors \cite{Radisavljevic:2011aa,Sarkar:2015aa,Dumcenco_2015,Dumcenco2015a,Mak2016,Zhang2018} and permits optical applications not attainable in the indirect band gap parent materials \cite{Splendiani2010,Bernardi_2013,Gutierrez2013,Das2014,Lopez-Sanchez:2013aa,Withers:2015aa,DiBartolomeo2018}. To realize this, large flakes of high quality materials are desirable and the presence of differently oriented domains is not a fundamental limitation, apart from the extended defects induced by the presence of domain boundaries. The exploitation of the valley \cite{Mak:2012aa,Zeng:2012aa,Cao:2012ab,Mak:2014aa} and spin degrees of freedom \cite{Xiao:2012ab,Xu:2014ac,Rycerz:2007aa,Xiao:2007aa,Zhang:2014ac}, on the other hand, requires a specific orientation of the material's unit cell and thus the absence of mirror domains. This is illustrated in Fig.~\ref{fig1}a, which shows the unit cell and electronic structure of SL TMDC mirror domains, illustrating the spin-reversal in the valence band maxima near $K$ and $-K$. For a simultaneous presence of two twin domains, the spin and valley polarization is lost on average and the observation of a valley Hall effect is prevented.

While van der Waals epitaxy of SL TMDCs on weakly interacting substrates such as sapphire\cite{Dumcenco_2015,Dumcenco2015a}, silicon oxide \cite{Zhang2017} and graphene \cite{Shi:2012aa} yields an angular distribution of domain orientations, highly crystalline films are achieved on h-BN by using very high growth temperature\cite{Fu2017}. On the other hand, growth on a more strongly coupling substrate results in two mirror domains aligned with the substrate lattice. A well-studied case is the epitaxial growth of SL MoS$_2$, the prototypical TMDC, on a Au(111) single crystal surface \cite{Miwa:2015aa}. The presence of two mirror domains is particularly evident in the initial stages of the growth when two types of triangular MoS$_2$ nano-islands are found, rotated by 180$^{\circ}$ with respect to each other \cite{Lauritsen:2007aa}. The simultaneous presence of mirror domains is detectable neither in the position of diffraction spots nor in the band structure obtained from angle-resolved photoemission (ARPES), at least in a non-spin resolved experiment. However, for a non-equal distribution of the two mirror domain areas, a finite average spin polarization or circular dichroism for excitations across the bands might still be detectable  \cite{Ulstrup:2017aa}.

Herein, we report on the growth of singly-oriented SL MoS$_2$ on Au(111) and measure the complete spin polarization of the states located near $K$ and $-K$ by means of spin-resolved ARPES.

\section{Results and Discussion}
The growth procedure used to synthesize SL MoS$_2$ 
differs from that reported earlier \cite{Sorensen_2014,Miwa:2015aa,Lauritsen:2007aa} since here the synthesis takes place at high temperature. In short, Mo was evaporated  in a background pressure of $2\times10^{-6}$ mbar of H$_2$S onto a clean Au(111) surface while the substrate was  kept at a temperature of $823$~K. 
These conditions were determined by following during the growth the behavior of the Mo and S core levels measured in real time with fast X-ray photoemission spectroscopy (XPS) as shown in Figure S1.  A full description of the sample preparation is given in the Methods section. 

\begin{figure*}[tbp]
\begin{center}
\includegraphics[width=0.8\textwidth]{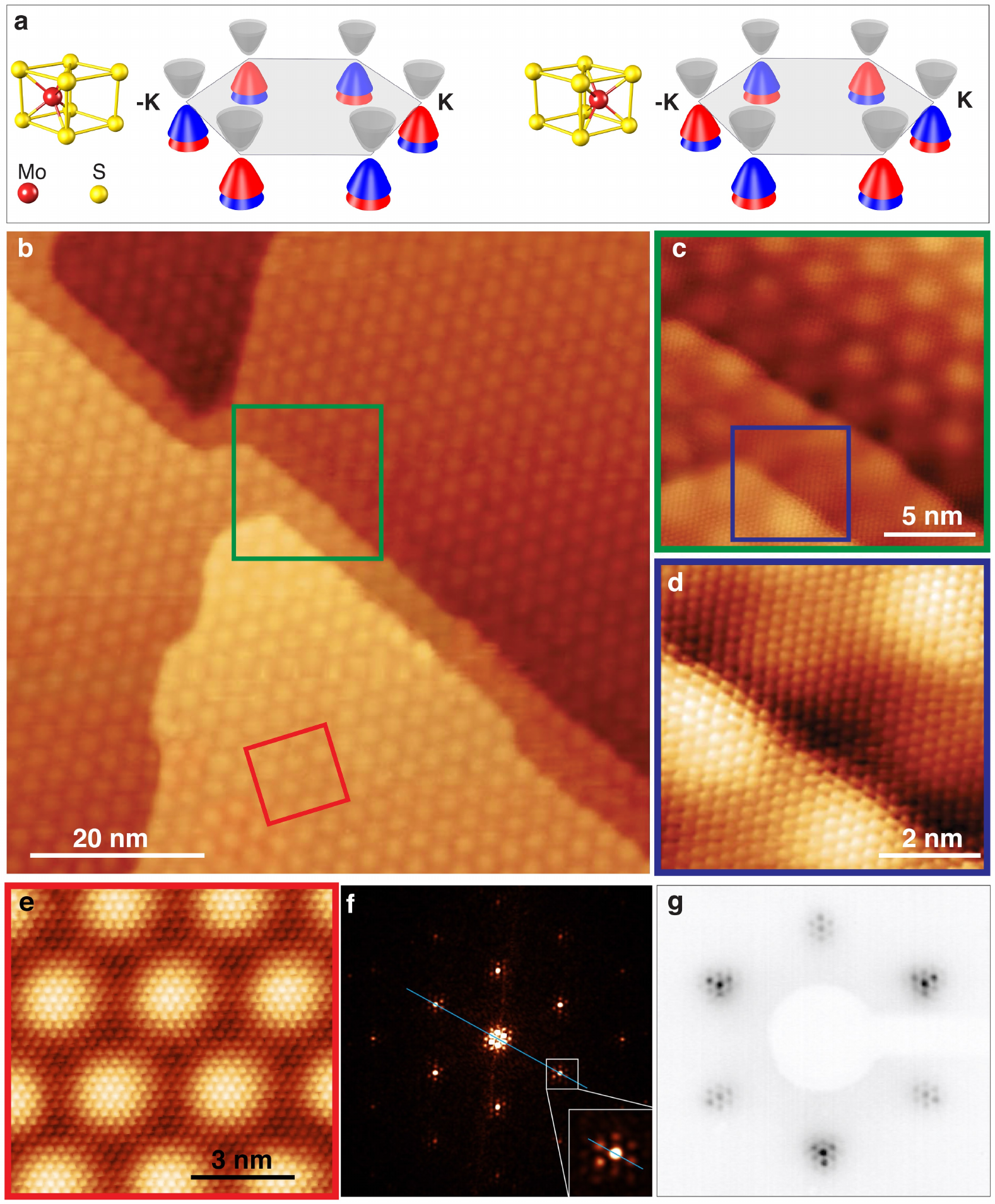}
\caption{STM and LEED characterization of single-orientation SL MoS$_2$. (a) Structure, Brillouin zone and schematic band structure for two mirror domain orientations of SL MoS$_2$. The colors of the split valence band maximum (blue/red) refer to the different spin orientation of these states. (b) STM topography acquired on a large area $V_T=0.525$~V, $I_T=1.04$~nA). (c) and (d) STM images acquired close to an atomic step of the Au substrate (image c: $V_T=0.525$~V, $I_T=1.05$~nA, image d: $V_T=0.525$~V, $I_T=1.05$~nA). The different areas probed are framed in different colors. (e) STM image ($V_T=0.525$~V, $I_T=0.89$~nA) acquired on a Au terrace (red frame) and (f) corresponding FFT analysis of the image shown in (e). A detail of the framed spot is magnified in the bottom inset. (g) LEED pattern ($E_p=$185~eV)}
  \label{fig1}
  \end{center}
\end{figure*}
The overall structural properties of the resulting layer were characterized by scanning tunneling microscopy (STM) and low-energy electron diffraction (LEED).  
The STM images (Figure 1b, c and d) show a small-scale hexagonal atomic structure due to the top sulfur layer of the S-Mo-S structure of MoS$_2$ together with a large scale moir\'e pattern due to the lattice mismatch between MoS$_2$ and Au(111). The moir\'{e} superlattice is well visible on the Au terraces and maintains its orientation across the Au atomic steps. The atomically-resolved images acquired in the green- and blue-framed regions (Figure \ref{fig1}c and d, respectively) show that the MoS$_2$ layer extends over the Au atomic steps with carpeting effect. The large area image (Figure \ref{fig1}b) evidences the lack of any domain boundaries or dislocations on the entire probed area ($\sim80\times80\:\text{nm}^2$). 

Fast-Fourier transform (FFT) analysis (figure \ref{fig1}f) was carried out on the representative atomically-resolved image measured on a Au terrace (Figure~\ref{fig1}e). The alignment of the FFT spots as indicated by the blue line in the figure shows that the MoS$_2$ layer is aligned along the direction of the moir\'{e} superstructure, and thus aligned along the crystallographic axes of the Au(111) substrate. This implies that only two orientations of the MoS$_2$ layer are possible, rotated by 180$^{\circ}$ with respect to each other. This finding is at variance with respect to the results reported in literature  for SL MoS$_2$ grown with the earlier synthesis method \cite{Sorensen_2014,Miwa:2015aa,Lauritsen:2007aa}, for which a misalignment angle of $0.45^{\circ}$ between the MoS$_2$ and the Au substrate was observed \cite{Sorensen_2014}. Moreover, by comparing this outcome with the LEED (Figure \ref{fig1}g) and SPA-LEED patterns (see below) we can deduce that the moir\'{e} superstructure is due to the $10\times10$ surface unit cell of MoS$_2$ over the $11\times11$ unit cell of Au(111).

\begin{figure*}[t]
\begin{center}
\includegraphics[width=0.8\textwidth]{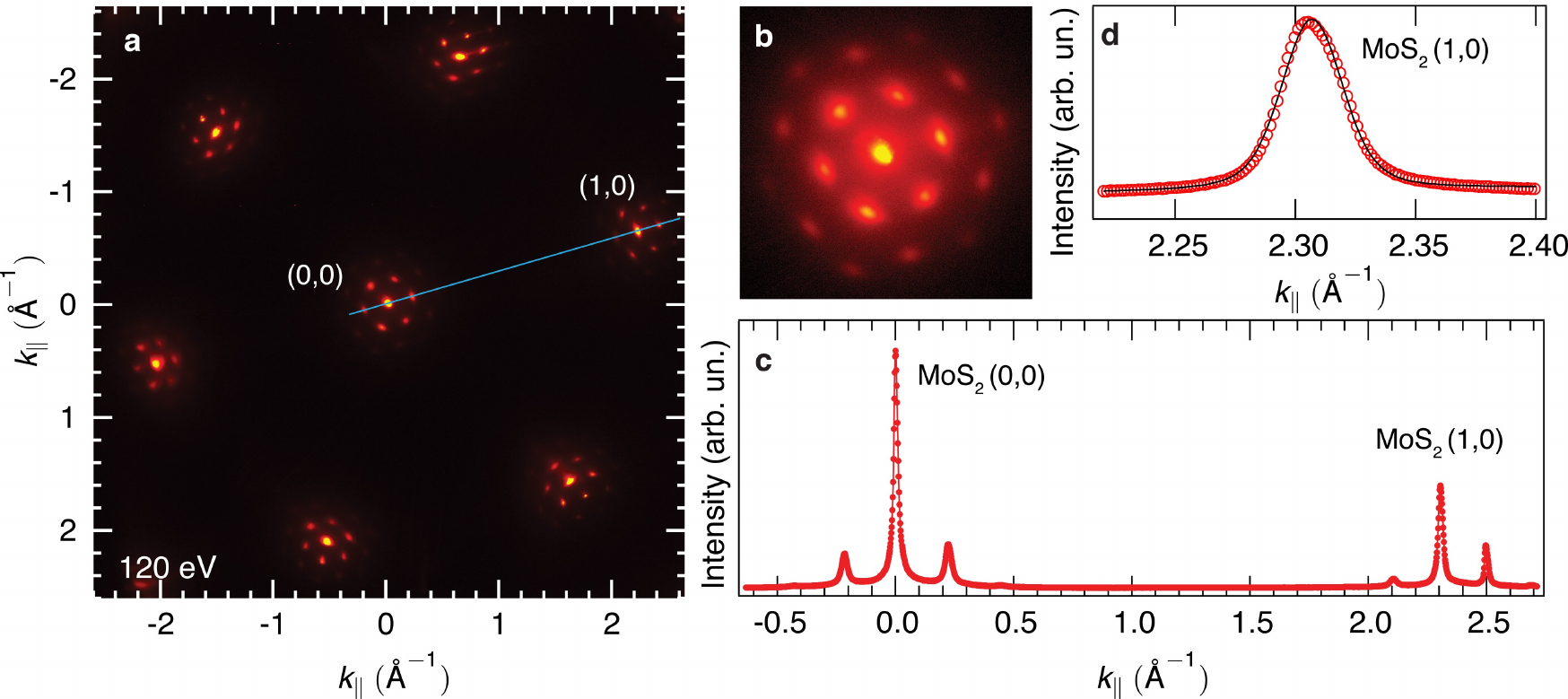}
\caption{SPA-LEED measurements on SL MoS$_2$. (a) Two-dimensional SPA-LEED pattern measured at a kinetic energy of 120~eV. (b) Detail acquired around the (0,0) diffraction spot. (c) Spot profile along the [10] direction (blue line shown in Figure \ref{SPA}a). (d) Detail of the (1,0) diffraction spot with profile analysis. }
\label{SPA}
\end{center}
\end{figure*}
To measure the average domain size of the MoS$_2$ layer we performed high-resolution \mbox{$k$-space} measurements by means of spot-profile analysis (SPA)-LEED. The two-dimensional pattern measured at a kinetic energy of 120 eV is shown in Figure~\ref{SPA}a. Besides the zeroth- and first-order spots, the image shows the appearance of additional diffraction spots due to the moir\'{e} superlattice. This can be appreciated in Figure~\ref{SPA}b, where the k-space has been probed around the $(0,0)$ diffraction beam. The spot profile along the $[10]$ direction is presented in Figure \ref{SPA}c. The panel in Figure \ref{SPA}d shows a detail of the (1,0) MoS$_2$ spot together with the best fit analysis obtained using a Voigt function. The Gaussian width is 0.0210~\AA$^{-1}$, while the Lorentzian full width at half maximum ($L_w$) is 0.0094~\AA$^{-1}$, corresponding to an average domain size of $1040\pm50$~\AA. Since this is comparable with the transfer width of the instrument, the average domain size could largely exceed this value. 

\begin{figure*}[t]
\begin{center}
\includegraphics[width=0.8\textwidth]{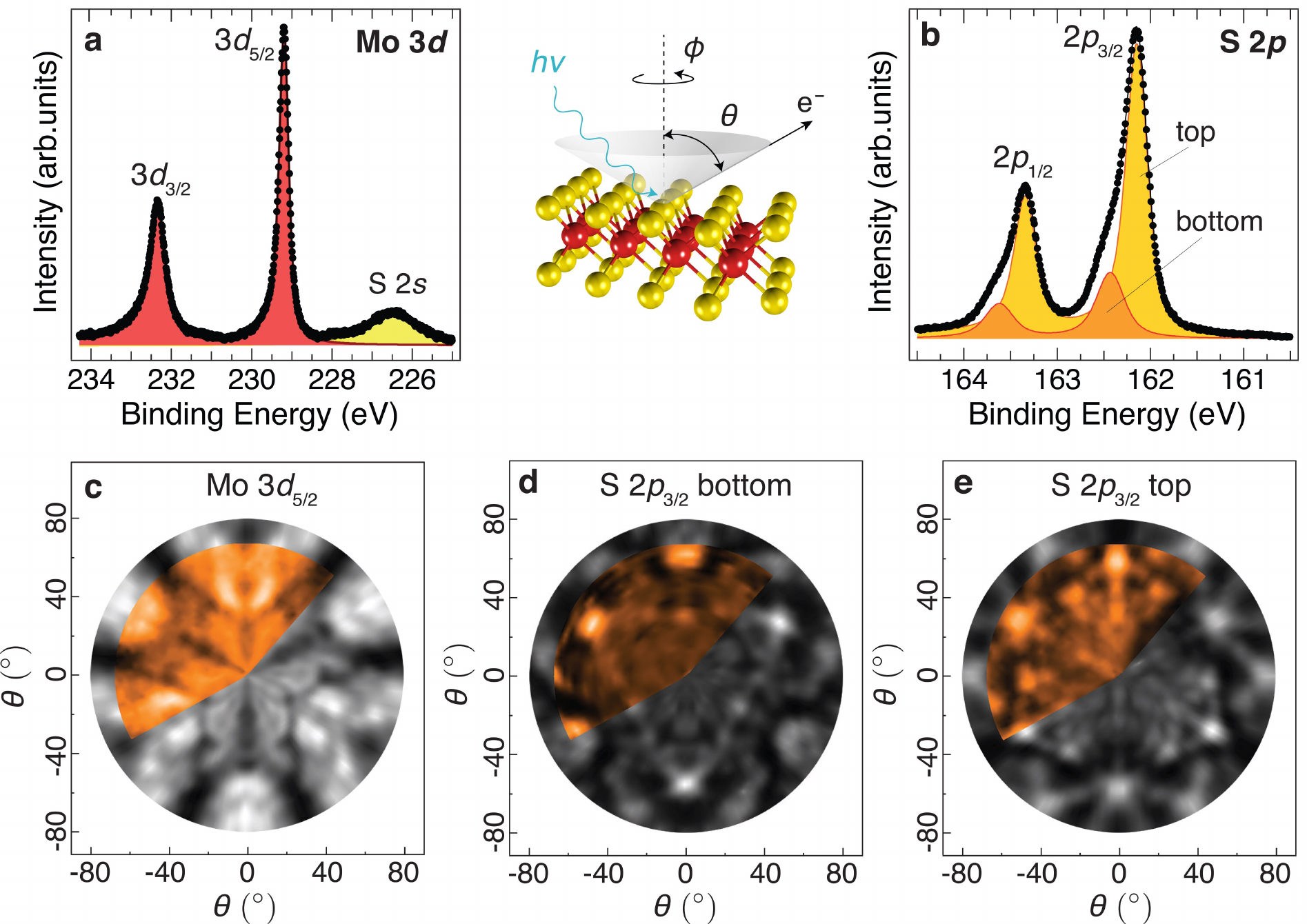}
\caption{XPS and XPD characterization of SL MoS$_2$. (a) Mo 3d core level spectrum taken at $h\nu$=360~eV (data points) with the resulting fit (line) and the fitted components (solid areas). (b) S 2$p$ core level spectrum taken at $h\nu$=260~eV (data points) with the resulting fit (line) and the fitted components (solid areas). The light and dark orange peaks correspond to the upper and lower sulfur layers, respectively. (c), (d), (e) Stereographic projections of the modulation function $\chi$ for Mo 3$d_{5/2}$, S 2$p_{3/2}$ bottom and S 2$p_{3/2}$ top taken at a photon energy of 360~eV, 560~eV and 270~eV, corresponding to a kinetic energy of 130~eV, 397~eV and 108~eV, respectively. The colored sectors are the experimental data and the greyscale disks are the XPD patterns simulated for a layer with a single orientation and with the structure determined as specified in the following. The sketch in the centre depicts the XPD experimental geometry, where $\theta$ and $\phi$ are the polar and azimuthal angle, respectively.}
\label{Fig3}
\end{center}
\end{figure*}
The high structural quality is also reflected in core level spectra obtained by XPS and shown in Figure \ref{Fig3}a and b for Mo 3$d$ and S 2$p$, respectively. The Mo 3$d$ spectrum can be fitted with a  doublet (red) with the Mo~3$d_{5/2}$ centered at 229.19~eV (spin-orbit splitting of 3.15~eV). The broad peak at  226.43~eV (yellow) is the S 2$s$ core level. The S 2$p$ core level shows two strong spin-orbit doublets \cite{Sorensen_2014}. The more intense peak at 162.15~eV (light orange) is assigned to the S 2$p_{3/2}$ core level of the upper sulfur layer (i.e. the layer towards vacuum) and the weaker at 162.44~eV (dark orange) to the layer towards the Au surface. The width of these components (table S1) and the absence of additional peaks related to sulfided species not converted into MoS$_2$ or to the atoms at the edges of the MoS$_2$ islands \cite{Bruix_2015} are indicative of the high quality of the layer, in accordance with the STM and LEED results. 
 
The single orientation of the MoS$_2$ layer can be ascertained in an x-ray photoelectron diffraction (XPD) experiment, as already successfully demonstrated for h-BN\cite{Orlando_2014}. This technique is based on emission-angle-dependent modulations of the core level photoemission intensity from the different atoms in the layer \cite{xpd_rfac}. The intensity modulations arise from the length difference between individual scattering pathways from the emitting atom to the detector and the coherent interference of the scattered waves. The XPD modulations are thus directly reflecting the local structural environment of the emitting atom. 

Figure \ref{Fig3}c, d and e show stereographic projections of the modulation function $\chi$ (see Methods) for Mo and the lower and upper S atoms, respectively. The colored part is the data and the greyscale part is a simulation for a layer with a single orientation (see below). The kinetic energy of the photoelectrons for the main XPS line in these experiments can be chosen by tuning the incoming photon energy; it was set high for the lower S atoms, favoring forward scattering processes from the Mo and S above, and low for the Mo and upper S atoms, favoring backscattering processes. All three diffraction patterns show a clear three-fold symmetry. Assuming a negligible influence of the underlying Au surface on the symmetry of the pattern, this already excludes the presence of equal areas of mirror domains, since these would give rise to a six-fold pattern.

\begin{figure*}[t]
\begin{center}
\includegraphics[width=0.8\textwidth]{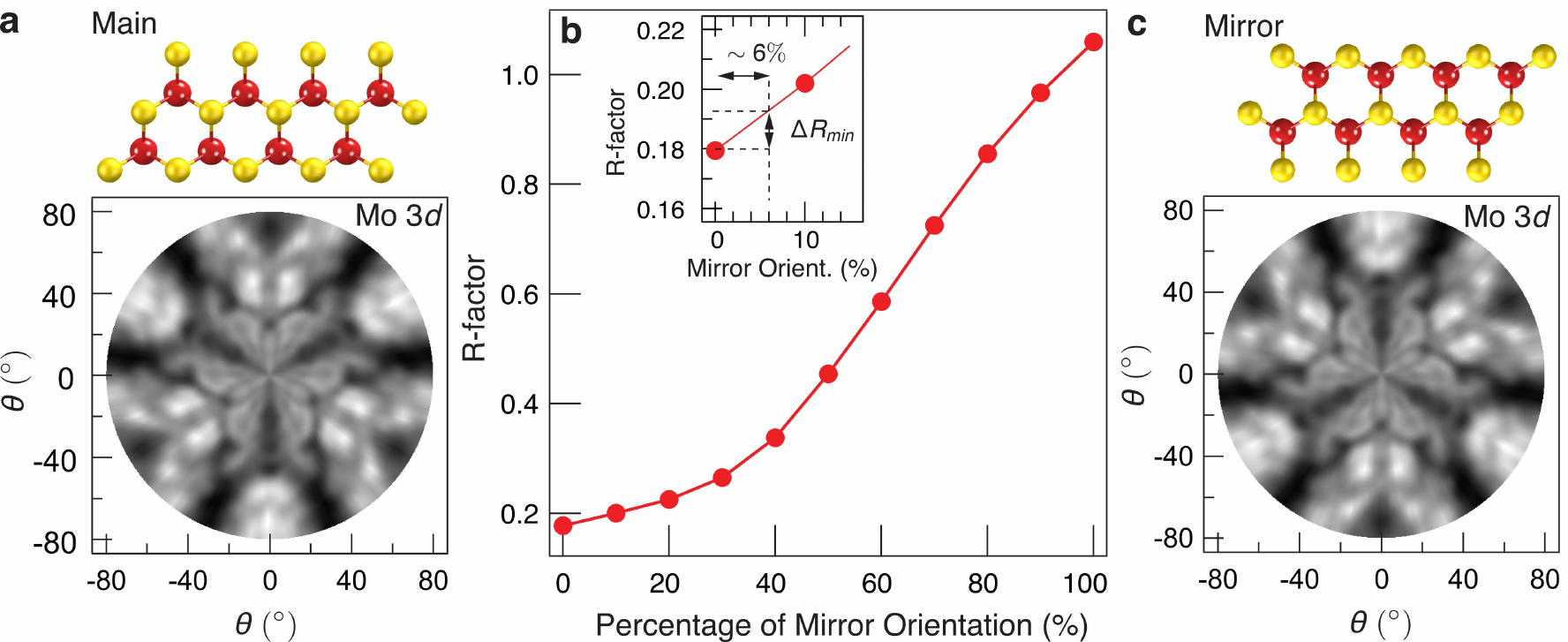}
\caption{MoS$_2$ layer orientation from XPD quantitative analysis. Atomic ball-model and corresponding simulated XPD pattern for the MoS$_2$ domains aligned along the main (a) and mirror (c) orientation. (b), $R$-factor behavior for the Mo 3$d_{5/2}$ diffraction pattern, as a function of the relative concentration of the mirror orientation, with a lattice parameter and S-S inter-plane distance of 3.17~\AA. The inset shows a magnification of the graph around the minimum of the $R$-factor, with the dashed lines indicating the confidence interval $\Delta R_{min}$.
} 
 \label{fig4}
 \end{center}
\end{figure*}
For further analysis, simulated diffraction patterns were calculated for the Mo~$3d_{5/2}$ of trial MoS$_2$ structures using the software package for Electron Diffraction in Atomic Clusters (EDAC)\cite{Garcia_2001}. In these simulations the underlying Au surface was totally neglected, which is appropriate because of the lack of a specific local adsorption configuration of MoS$_2$ on the substrate due to the lattice mismatch with Au(111).
The orientation and the structure of SL MoS$_2$ were determined by minimizing the Reliability ($R$) factor as a function of the abundance of the two possible orientations observed with STM, of the lattice parameter and of the  S-S inter-plane separation. (see Supplementary Information for details). 
The total intensity $I_{\text{tot}}$ in the simulated diffraction patterns for different admixtures of the two mirror orientations can be expressed as 
\begin{equation}
I_{\text{tot}}=aI_{0}+bI_{\text{mir}}\qquad(b=1-a)
\end{equation}
being $I_0$ the contribution to the  XPD pattern sourcing from the main orientation (Figure \ref{fig4}a) and $I_{\text{mir}}$ the contribution from the mirror orientation (Figure \ref{fig4}c). 
 As displayed in Figure \ref{fig4}b, the R-factor shows a minimum when only the main orientation is present, while the agreement is worsened for any admixture of the two orientations. We estimated the maximum amount of the  mirror orientation consistent with our data by calculating the confidence interval displayed in the inset of Figure \ref{Fig3}b (see Supplementary Information for details). Based on this analysis, we can infer that the fraction of the mirror orientation in the MoS$_2$ layer does not exceed~$\sim$6~\%. 
These outcomes are consistent with the results stemming from the LEED pattern  in Figure~\ref{fig1}g showing a clear three-fold symmetry, although these observations  alone would not be sufficient to establish the domain orientation, which can be conclusively determined by XPD.
The single orientation growth on Au(111) is likely due to the symmetry breaking originating from the substrate. While the first atomic layer of Au(111) has a six-fold symmetry and would permit both MoS$_2$ orientations, when the deeper Au layers are considered the crystal symmetry results to be three-fold. This apparently tips the balance between the two possible aligned orientations towards a single one. (See Supplementary Information for the details on the stacking registry between MoS$_2$ and the Au(111) substrate)

\begin{figure*}[t]
\begin{center}
\includegraphics[width=0.8\textwidth]{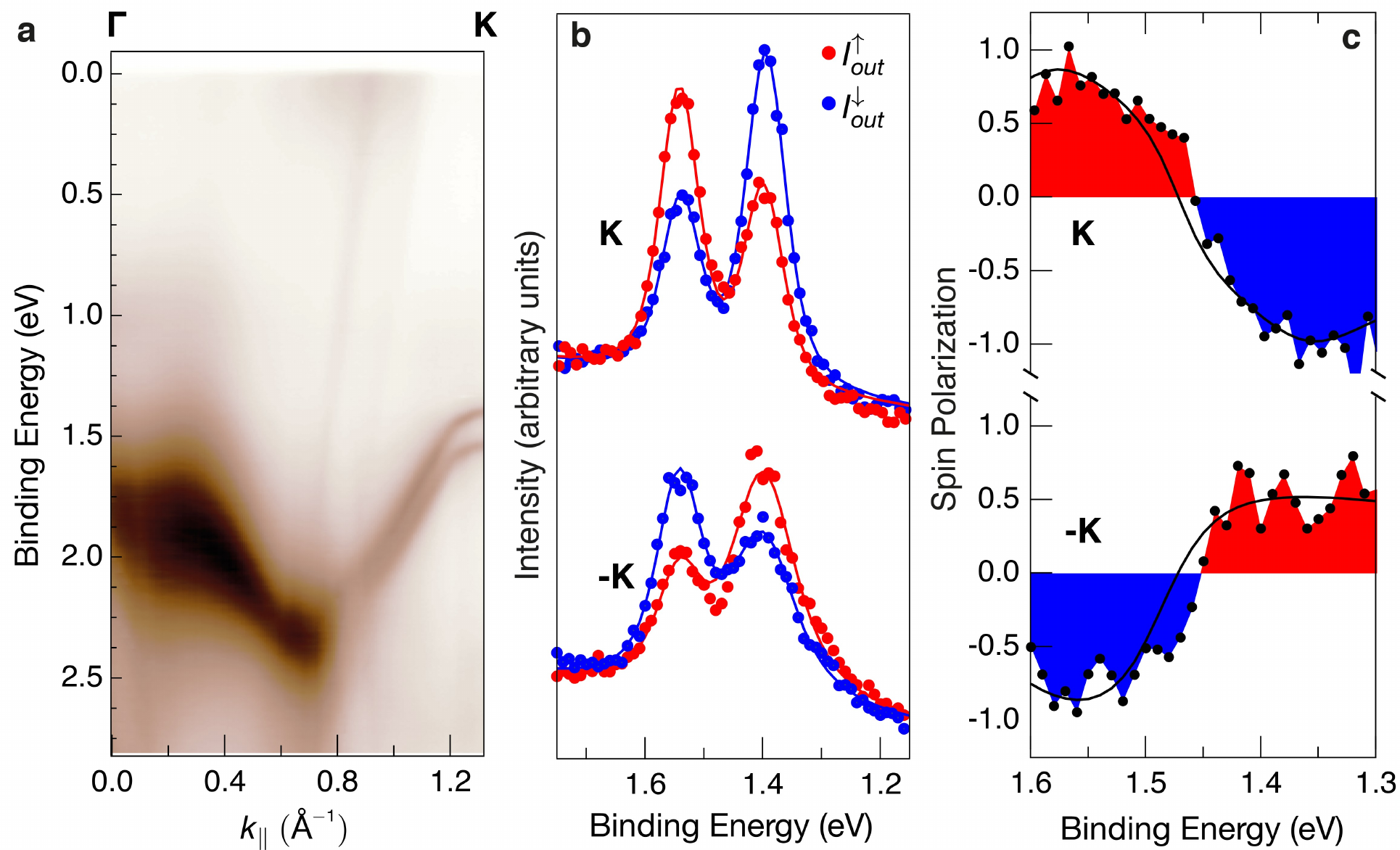}
\caption{Spin polarization of singly-oriented SL MoS$_2$. (a) Angle-resolved photoemission intensity ($h\nu=25\:\text{eV}$) along the $\Gamma-K$ direction of  SL MoS$_2$  Brillouin zone. (b) Out-of plane spin-resolved energy distribution curves at $K$ and $-K$ points ($h\nu=30\:\text{eV}$). Red and blue colors mark spin-up and -down signals, respectively. Raw data is shown without a correction for the spin-sensitivity (Sherman function) of the detector. Solid lines are Voigt fittings to the experimental data marked with circles. (c) Background-subtracted out-of plane spin polarization at $K$ and $-K$ points. Solid lines mark the spin polarization calculated from the fits, taking the Sherman function into account.}
 \label{fig5}
 \end{center}
\end{figure*}
Having established the presence of a single domain orientation by analyzing the geometric structure of the layer, we proceed by demonstrating its effect on the electronic structure. Figure \ref{fig5}a shows the dispersion of the MoS$_2$ bands measured by ARPES. The valence band maximum at $K$ is clearly visible, including the spin-orbit splitting of the state. As pointed out above, the band structure observed in an ARPES measurement without spin resolution is not substantially affected by the presence of mirror domains. Many small domains would merely lead to a spectral broadening due to defect scattering. 
While the data in Figure \ref{fig5}a are similar to previous findings for this system \cite{Miwa:2015aa,Bruix:2016aa}, the linewidth of the states near $K$ is substantially smaller (51 and 70~meV for the upper and lower band, respectively) than what reported earlier \cite{Miwa:2015aa}, indicating a higher quality of the layer. The ARPES results show that the system has no detectable contributions from a second layer, as this would be observed as a second band near $\Gamma$ (see Supplementary for a detailed discussion about this issue).\cite{Zhang:2014aa,Miwa:2015ab}

The single orientation character of the layer can be expected to result in a complete spin polarization of the bands near $K$ and $-K$ and Figure \ref{fig5}b and c show an experimental demonstration of this using spin-resolved photoemission spectroscopy. We find an out-of plane spin polarization of 86$\pm$14\%, which is opposite for $K$ and $-K$. In case of multiple-domain crystals, we expect the measured signal to be a mixture of contributions from oppositely spin-polarized  $K$ and $-K$ valleys, leading to a decreased value of spin polarization. Here, on the contrary, we measure a high magnitude of spin polarization, which further confirms the single domain orientation of the MoS$_2$ monolayer.

Previous spin-resolved ARPES experiments on bulk TMDCs have revealed a surprising complexity in the observed spin texture. Naively, inversion symmetry should lead to no observable spin polarization from the 2H structural polymorph while the 3R polymorph could give rise to a spin-polarized signal. The latter was indeed found for 3R MoS$_2$ \cite{Suzuki:2014aa} while, surprisingly, also 2H WSe$_2$ revealed strongly spin-polarized bands, essentially due to a combination of local symmetry breaking and surface sensitivity in photoemission \cite{Riley:2014aa}. Later, it was shown that the observed spin polarization from 2H MoS$_2$ could even be switched by excitation with circularly polarized light of different handedness \cite{Razzoli:2017aa}. In simple terms, this effect is based on the coupling of the light to different valleys in the band structure and the locking of valley and layer degrees of freedom in the 2H structure. 
Such a switching effect, and more in general a strong dependence on photon energy and light polarization of the spin-polarization, should not be observable in the case of a single-orientation single layer\cite{Razzoli:2017aa}. 
This is confirmed by results for SL MoSe$_2$ grown on bilayer graphene~\cite{Sung-Kwan:2016}: the spin polarization was not found to change significantly with photon energy, even if it is much smaller than the value reported here, due to the presence of mirror domains. However, due to the complexity of the spin-polarized photoemission process, changes of the observed polarization due to e.g. final state effects cannot be excluded. We thus emphasize that the observed spin polarization in our case is consistent with the presence of a single domain, but that our analysis of the domain distribution rests not on this but on the structural analysis based on XPD.

\section{Conclusions}
In summary, we have presented a synthesis method to produce high quality, singly-oriented SL MoS$_2$. We used a multi-method approach  to determine  the structural properties of the layer and we measured for the first time the complete spin polarization of the states near $K $and $-K$ of SL MoS$_2$ by spin-resolved ARPES. 

The synthesis method  outlined here may represent a breakthrough for the large scale production of high-quality MoS$_2$ monolayers with a low number of dislocation defects. The availability of the singly-oriented MoS$_2$ monolayers obtained with this protocol may boost the research on the spin-valley degree of freedom in two-dimensional materials and could be the key to realize mass-produced devices based on the valleytronics concept.
This growth protocol could potentially be applied also for the synthesis of high-quality singly-oriented WS$_2$ or MoSe$_2$ monolayers on Au(111). Guided by the developments in graphene synthesis, one could expect that this method can be employed on other substrates or that destruction-free transfer mechanisms for large areas can be devised. 

\section{Experimental Methods}

The growth of MoS$_2$ samples, the LEED, the high-resolution XPS and the XPD experiments were carried out at the SuperESCA beamline of the Elettra synchrotron radiation facility in Trieste (Italy) \cite{Baraldi2003}. The UHV experimental chamber is equipped with a Phoibos hemispherical electron energy analyzer (150 mm mean radius), implemented with a home-made delay line detector. The experimental chamber is equipped also with a rear-view LEED system. The Au single crystal was fixed on a Ta plate and the sample temperature was measured by two thermocouples spot-welded very close to it. The sample holder was mounted on a 5 degrees of freedom ($x$, $y$, $z$, $\theta$, $\phi$) manipulator.
 
 The Au(111) surface was prepared by repeated cycles of Ar$^+$ sputtering followed by annealing up to 920 K for 10 min. The heating and cooling rate was 1 K/s. After the cleaning procedure, the sample cleanliness was checked with XPS, which did not detect any trace of contaminants within the detection limit of 0.1\% of a monolayer (ML) where 1 ML corresponds to the surface atomic density of the Au(111) surface. The long-range order was verified by acquiring the LEED pattern on the freshly prepared sample, which showed the extra spots of the herringbone reconstruction.
 
MoS$_2$ monolayers were grown by dosing molybdenum from a home-built evaporator, consisting of a Mo filament annealed through direct current heating, while keeping the Au substrate at 823 K and dosing H$_2$S (nominal purity 99.8\%) through a leak valve at background pressure of $2\times10^{-6}$ mbar. 
The Mo deposition rate was estimated by means of a quartz microbalance  and amounted to $\sim$ $5\times10^{-3}$~ML/minute. Therefore, the total amount of Mo deposited in 8000~s was $\sim$~0.67~ML. From the attenuation of the surface component of the Au 4$f$ core level due to the presence of the MoS$_2$ layer, we estimated a MoS$_2$ coverage of 0.65~ML (1~ML corresponds here to one layer of MoS$_2$ covering the whole surface). 
High-resolution S~$2p$ and Mo~$3d$ core level spectra were measured at room temperature on the as-grown MoS$_2$ monolayer, using photon energies of 260 eV and 360 eV, respectively. The overall energy resolution was better than 50 meV. The surface normal, the incident beam, and the electron emission direction were all in the horizontal plane, with the angle between the photon beam and the electron energy analyser fixed at $70^{\circ}$. The high resolution core level spectra were acquired at normal electron emission.  

SPA-LEED measurements were carried out at the Surface Science Laboratory of Elettra Sincrotrone Trieste using a commercial Omicron SPA-LEED \cite{Lizzit2009}. The transfer width was better than 1000~\AA. The instrument was used to acquire two-dimensional diffraction patterns at fixed energy as well as to measure one-dimensional high-resolution profiles along specific reciprocal space directions. The line-profile of the diffraction spots were modeled with a Voigt function \cite{Locatelli2010}. The Gaussian broadening accounts for the finite coherence length of the primary electron beam and for the corrugation of the substrate. The Lorentzian contribution is connected with the size of the MoS$_2$ domains on the surface through the formula:
\begin{equation}\label{TerrD}
\frac{L_w}{a^*}=\frac{a}{D}
\end{equation}
where $L_{w}$ is the full-width at half maximum of the Lorentzian component, $a^*$ and $a$ are the reciprocal and real lattice vectors of MoS$_2$, respectively, and $D$ is the average width of the crystalline domains.  

XPD patterns for Mo $3d$ and S $2p$ core levels were acquired with different photon energies ($h\nu$) in order to change the corresponding electron kinetic energy (KE) to enhance forward and backscattering conditions.
Specifically, Mo~$3d$ was acquired with $h\nu=360$~eV corresponding to electron KE of $\sim$130~eV.
The S~$2p$ XPD pattern from the top S layer was acquired at $h\nu=270$~eV (electron KE of $\sim$108~eV) to enhance backscattering conditions, while the pattern from the bottom S layer was measured with $h\nu=560$~eV (electron KE of $\sim$397~eV) to enhance forward scattering conditions. 
At each energy more than 1000 spectra were measured for different polar ($\theta$) and azimuthal ($\phi$) angles. For each of these spectra, the peak fit analysis was performed and the intensity $I(\theta$, $\phi)$ of each component resulting from the fit, \textit{i.e.} the area under the photoemission line, was extracted. Each XPD pattern was measured over an azimuthal sector of $160^{\circ}$, from normal ($\theta=0^{\circ}$) to grazing emission ($\theta=70^{\circ}$), as shown in the sketch in the centre of Figure 3 of the main text. 
The resulting XPD patterns are the stereographic projection of the modulation function $\chi$, which was obtained from the peak intensity for each polar emission angle as 
\begin{equation}
{\chi} = \frac{I(\theta, \phi) - I_0(\theta)}{I_0(\theta)}
\end{equation}
 where $I_0$($\theta$) is the average intensity for each azimuthal scan. The agreement between the simulations and the experimental results was quantified by computing the reliability factor (R),
\begin{equation}R = \frac{\sum_{i} (\chi_{\text{exp}, i} - \chi_{\text{sim}, i})^2}{\sum_i({\chi^2}_{\text{exp}, i} + {\chi^2}_{\text{sim}, i})}\end{equation} where  $\chi_{\text{sim}, i}$ and  $\chi_{\text{exp}, i}$ are the calculated and the experimental modulation functions for each data point $i$. The conformation of the MoS$_2$ layer was determined by minimizing the R-factor upon variation of the structural parameters employed in the simulations  \cite{xpd_rfac}.
Following the determination of the minimum $R-$factor, a confidence interval for the result was estimated by using the approach inspired by the common practice in LEED \cite{Pendry:1980aa}. The variance of the  minimum $R$-factor $R_{min}$ is calculated by 
\begin{equation}
\Delta R_{min} = \sqrt{2/N} R_{min},
\end{equation}
where $N$ is the number of well-resolved peaks in a LEED $I/V$ curve. Here we take $N=350$, which is the approximate number of peaks in the 50 azimuthal scans acquired at different polar emission angles. Consequently, having $R_{min}=0.18$ we find $\Delta R_{min}=0.0136$. 

ARPES experiments were carried out at the SGM-3 beamline of the synchrotron radiation facility ASTRID2 in Aarhus \cite{Hoffmann:2004aa}. The energy and angular resolution were better than 30~meV and 0.2$^{\circ}$, respectively. The sample temperature was $\sim$ 30~K. The sample was transferred to Aarhus in air. After inserting it into the ultrahigh vacuum system, it was annealed to 770~K to remove adsorbed impurities.  

Spin-resolved ARPES measurements were taken at the APE beamline of Elettra Sincrotrone Trieste, Italy \cite{Bigi2017}. The experimental chamber is equipped with a VG-Scienta DA30 analyzer and two very low energy electron diffraction (VLEED) spin polarimeters. Measurements were taken with a photon energy of 30~eV and p-polarized light, with the light incidence direction kept fixed at 45$^{\circ}$ with respect to the electron energy analyzer normal detection direction.  The energy and angular resolution were better than 50~meV and 0.75$^{\circ}$, respectively. Samples were transferred into the chamber in air and subsequently annealed up to $\sim$~800~K. Measurements were taken at about 80~K.
Spin polarization $P_{i}$ was determined from spin-resolved energy dispersion curves (EDCs)~$I_{i}^{\uparrow,\downarrow}$:
\begin{equation}
P_{i}=\frac{I_{i}^{\uparrow}-I_{i}^{\downarrow}}{S(I_{i}^{\uparrow}+I_{i}^{\downarrow})},
\end{equation}
where $i=x,y,z$ denotes the spin quantization axis in the reference frame of the detection and $S=0.3$ is the Sherman function of the detector. $I_{i}^{\uparrow,\downarrow}$ were corrected by a relative efficiency calibration and fitted with Gaussian-broadened Lorentzian (Voigt) peaks. The background contribution consisting of spin-unpolarized tails of lower-lying Au states was taken into account for the calculated $P_{i}$ spectra. Quantitative spin polarization magnitudes were determined from the area ratio of the fitted peaks. $P_{i}$ were transformed into the sample's reference frame by applying an Euler's rotation matrix. 

STM measurements were carried out at the CoSMoS facility at Elettra Sincrotrone Trieste. The images were acquired at room temperature with a SPECS STM 150 Aarhus instrument equipped with a W tip. The samples were transferred through air from the growth chamber to the STM chamber, where they were subsequently annealed up to ca. 800~K for 30 min. 

The size of the surface areas probed varied between the different experimental techniques. Light spots sizes for the synchrotron radiation experiments were typically in the order of more than 100~$\mu$m, the electron beam size in SPA-LEED was about 100~$\mu$m while for LEED was several hundreds micrometers.  Most importantly, no spatial inhomogeneity or presence of a mirror domain was noted when probing different areas of the Au(111) crystal surface (several mm in diameter) with different techniques.

We thank David Abergel and Albert Bruix for stimulating discussions. This work was supported by the Danish Council for Independent Research, Natural Sciences under the Sapere Aude program (Grant No. DFF-4002-00029) and by VILLUM FONDEN via the Centre of Excellence for Dirac Materials (Grant No. 11744). Affiliation with the Center for Integrated Materials Research (iMAT) at Aarhus University is gratefully acknowledged.

\section{Supplementary Information}
\subsection{Tuning of the Growth Parameters by Fast XPS}
\begin{figure*}[!h]
 	\includegraphics[width=0.8\textwidth]{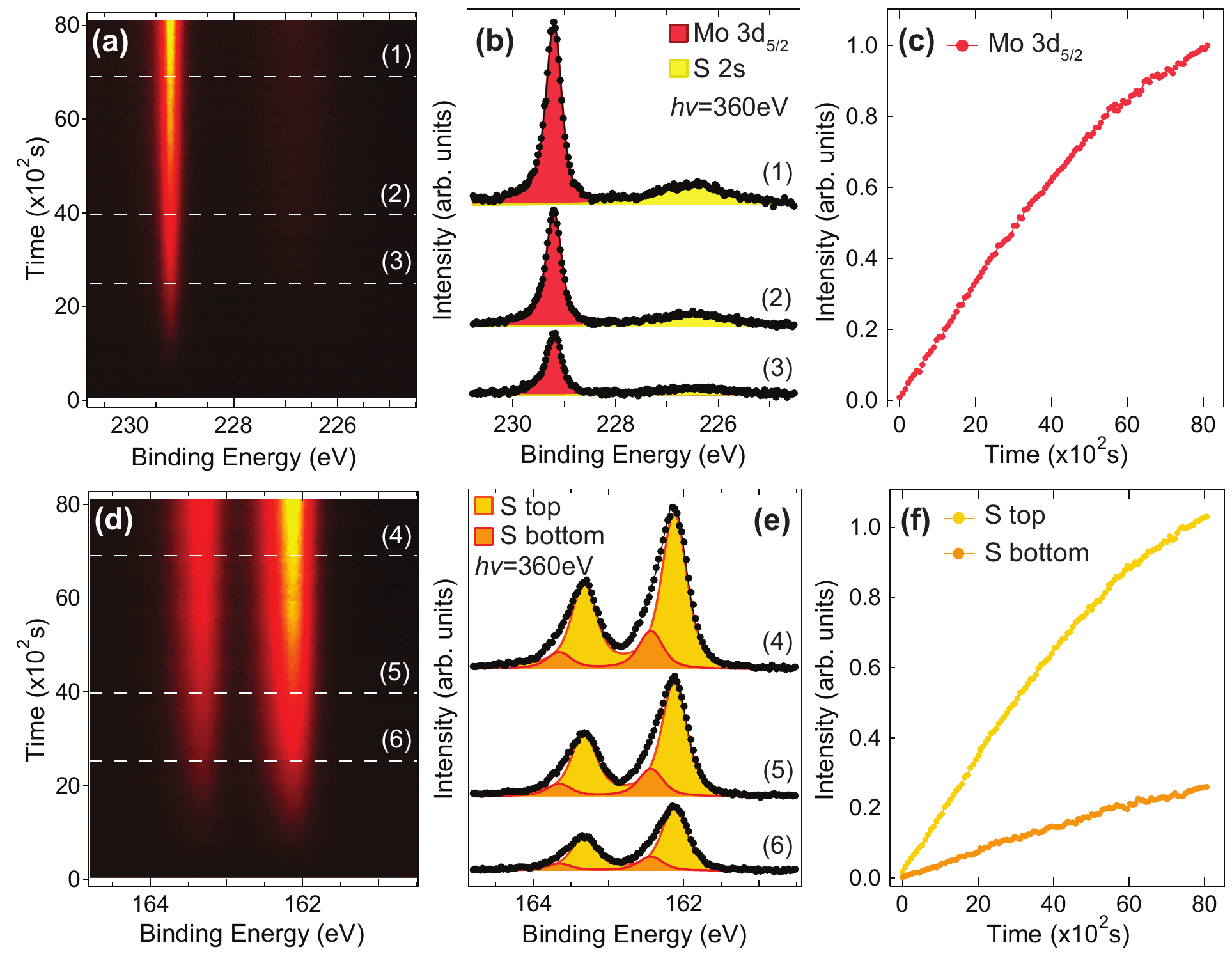}
 	\caption{Fast XPS during single layer MoS$_2$ growth. (a) and (d) show the intensity plot of the sequence of fast-XPS spectra of  Mo~$3d$ and S~$2p$ core levels, respectively, acquired simultaneously at 360 eV photon energy while growing MoS$_2$. Each spectrum was measured in $\sim$10~s. (b) Mo~$3d$ and (e) S~$2p$ core level spectra together with the spectral contributions resulting from the peak fit analysis acquired at different stages of the growth, as indicated by the dashed lines in (a) and (d), respectively. (c)  and (f) display the photoemission intensities obtained from the fit of the fast-XPS spectra, showing the evolution of the Mo 3$d_{5/2}$ and S 2$p_{3/2}$ for top and bottom sulfur, respectively.}
 	\label{fig1}
 \end{figure*}

The careful tuning of the growth parameters was achieved by following the real time evolution of the MoS$_2$ layer during the growth by means of fast-x-ray photoelectron spectroscopy (XPS) of the Mo~$3d$ and S~$2p$ core levels (Figure \ref{fig1}). In this way it was possible to avoid the growth of Mo clusters and partially sulfided species (core level peaks appearing at lower binding energy compared to those of MoS$_2$), which did not convert into MoS$_2$ even after prolonged annealing in H$_2$S atmosphere without dosing Mo. 
As one can see in Figure S1(c) and (f), the MoS$_2$ growth rate is constant until $\sim$2000 s and then it starts to decrease. This means that not all the Mo atoms react to form MoS$_2$ when the surface coverage increases, because of the partial desorption, at high temperature, of the Mo atoms impinging on the already formed MoS$_2$ layer.

\subsection{Core Level Spectra Lineshape}
\begin{table}[h]
 	\centering
 	
 	\label{table1}
 	\begin{tabular}{cccc}
 		\hline
 		{Mo $3d$ (\textit{hv}=360eV)}  & {L (eV)} & {$\alpha$} & {G (eV)}\\
 		
 		3$d_{5/2}$ & 0.20 & 0.06 & 0.14\\
 		3$d_{3/2}$ & 0.43 & 0.06 & 0.14\\

 		\hline
 		{S $2p$ (\textit{hv}=260eV)}  & {L (eV)} & {$\alpha$} & {G (eV)}\\
 		
 		2$p_{3/2}$ top (bottom) & 0.18 (0.23) & 0.05 (0.02) & 0.16 (0.21)\\
 		2$p_{1/2}$ top (bottom) & 0.18 (0.23) & 0.05 (0.02) & 0.16 (0.21)\\
 		\hline
		 		
 	\end{tabular}
	\caption{Doniach-\v{S}unji\'c line shape parameters for the different components of the Mo~$3d$ and S~$2p$ core levels acquired at room temperature with 360 eV and 260 eV photon energy, respectively. $L$ is the Lorentzian width which is related to the core-hole lifetime, $\alpha$ is the asymmetry parameter due to the electron-hole pairs creation near the Fermi level and G is the Gaussian width due to the instrumental resolution plus thermal and inhomogeneous broadening. }
 \end{table}
The core level spectra were fitted using a Doniach-\v{S}unji\'c line profile \cite{Doniach_1970} convoluted with a Gaussian broadening and a linear background. All binding energies presented in this work are referenced to the Fermi level of the Au substrate measured under the same experimental conditions. The fitting parameters for Mo~$3d$ and S~$2p$ core levels, for both top and bottom S, are listed in Table S1. The experimental error on the binding energy position of the peaks is $\pm10$~meV. 

\subsection{MoS$_2$ Monolayer vs Bilayer}
\begin{figure*}[t]
\begin{center}
\includegraphics[width=0.5\textwidth]{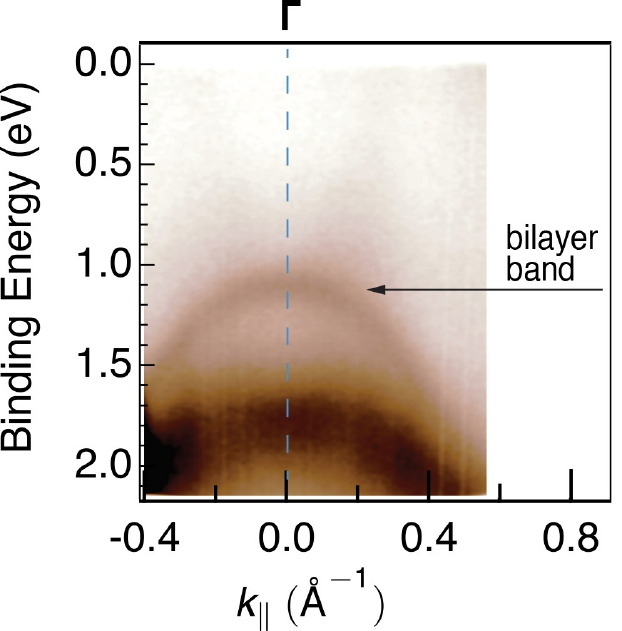}
\caption{ARPES of bilayer MoS$_2$. ARPES map acquired on a sample with bilayer regions. The spectra were measured around the $\Gamma$ point of the MoS$_2$ Brillouin zone with a photon energy of 25~eV. The sample temperature was 60~K. }
\label{ARPESbi}
\end{center}
\end{figure*}

It is possible to ascertain the presence of regions on the Au surface covered by bilayer MoS$_2$ by means of angle-resolved photoemission spectroscopy (ARPES), and thus discriminate samples where only monolayer is present. As earlier reported \cite{Miwa:2015aa,Miwa:2015ab,Gronborg:2015aa}, a second band at lower binding energy (BE) is visible at $\Gamma$ for MoS$_2$ bilayer, while it appears to be completely absent for monolayer MoS$_2$. 
In order to highlight these differences, we have grown a sample with a coverage exceeding 1~ML of MoS$_2$, where bilayer areas are likely to be found \cite{Gronborg:2015aa}. An ARPES map acquired in the vicinity of the $\Gamma$ point of the Brillouin zone is shown in Figure \ref{ARPESbi}. We can clearly see the appearance of a dispersing band centred at $\Gamma$ (pinpointed by an arrow) with the band maximum at about BE=1.1~eV. This feature is not present in the sample we employed for measurements described in the main paper (Figure 5), where the total MoS$_2$ coverage was 0.65~ML. Hence, we can conclude that no apparent bilayer contamination is present for the monolayer MoS$_2$ samples we have studied. STM analysis of the probed regions corroborated this conclusion, indicating that the possible presence of MoS$_2$ bilayers would be limited to less than 1\% of the sample surface.

\subsection{Stacking of MoS$_2$ on the Au substrate}
\begin{figure}[t]
\begin{center}
\includegraphics[width=0.5\textwidth]{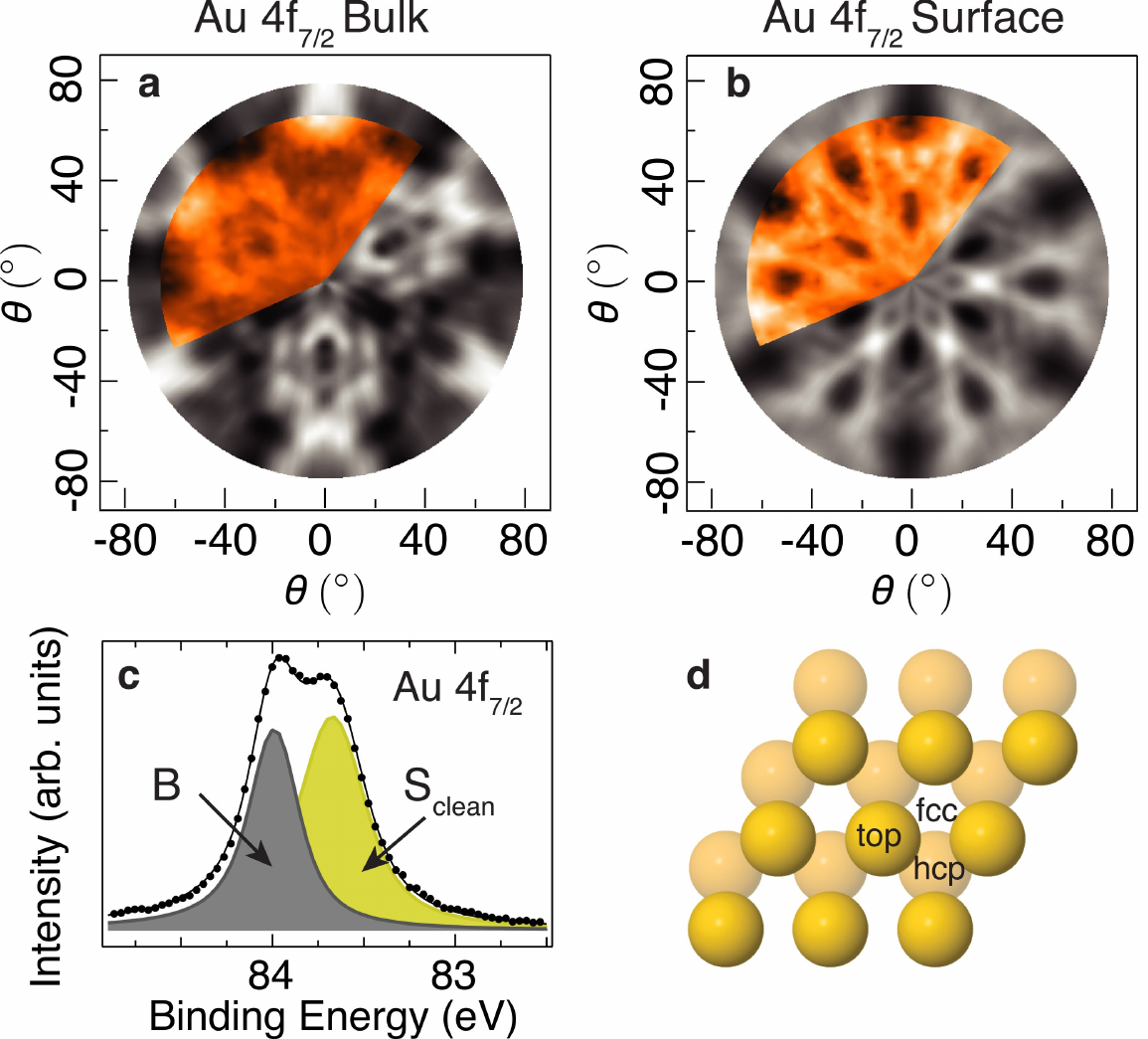}
\caption{Au $4f_{7/2}$ XPD patterns acquired at $\sim115$~eV photoelectron kinetic energy for (a) bulk (R-factor=0.33) and (b) surface (R-factor=0.33) components. The experimental patterns (colored) are compared with multiple scattering simulations (grey) and the orientation of the Au(111) crystal is identified as shown in (d). (c) Au $4f_{7/2}$ XPS spectrum acquired of the clean Au(111) sample at 200~eV photon energy, showing the bulk (B) and clean surface ($S_\mathrm{clean}$) components.}
\label{XPDAu}
\end{center}
\end{figure}

In order to determine the orientation of the Au(111) substrate, needed to identify the relative orientation of the MoS$_2$ layer, we performed XPD measurements of the Au $4f_{7/2}$ core level for the clean sample (Figure \ref{XPDAu}c). Figure \ref{XPDAu}a and b show the XPD patterns corresponding to the bulk (B) and clean surface ($S_\mathrm{clean}$) components (colored), respectively, measured at 200 eV photon energy (photoelectron kinetic energy $\sim115$~eV). The XPD pattern of the bulk component shows the expected three-fold symmetry of the \textit{fcc} crystal stacking, while the herringbone reconstruction presents an almost six-fold symmetry of the pattern from the Au $4f_{7/2}$ surface component. For this reason the XPD pattern of the bulk peak provides the orientation of the Au substrate. The herringbone reconstruction was simulated by compressing the surface unit cell in the $\langle-110\rangle$ direction by 4.5\%, averaging over the three $60^{\circ}$ rotated domains, while three layers below the surface were considered as the bulk. The experimental data show a good agreement (R-factor=0.33) with the simulations and the resulting orientation of the gold substrate was identified as shown in figure Figure \ref{XPDAu}d.

\begin{figure}[t]
\begin{center}
\includegraphics[width=0.7\textwidth]{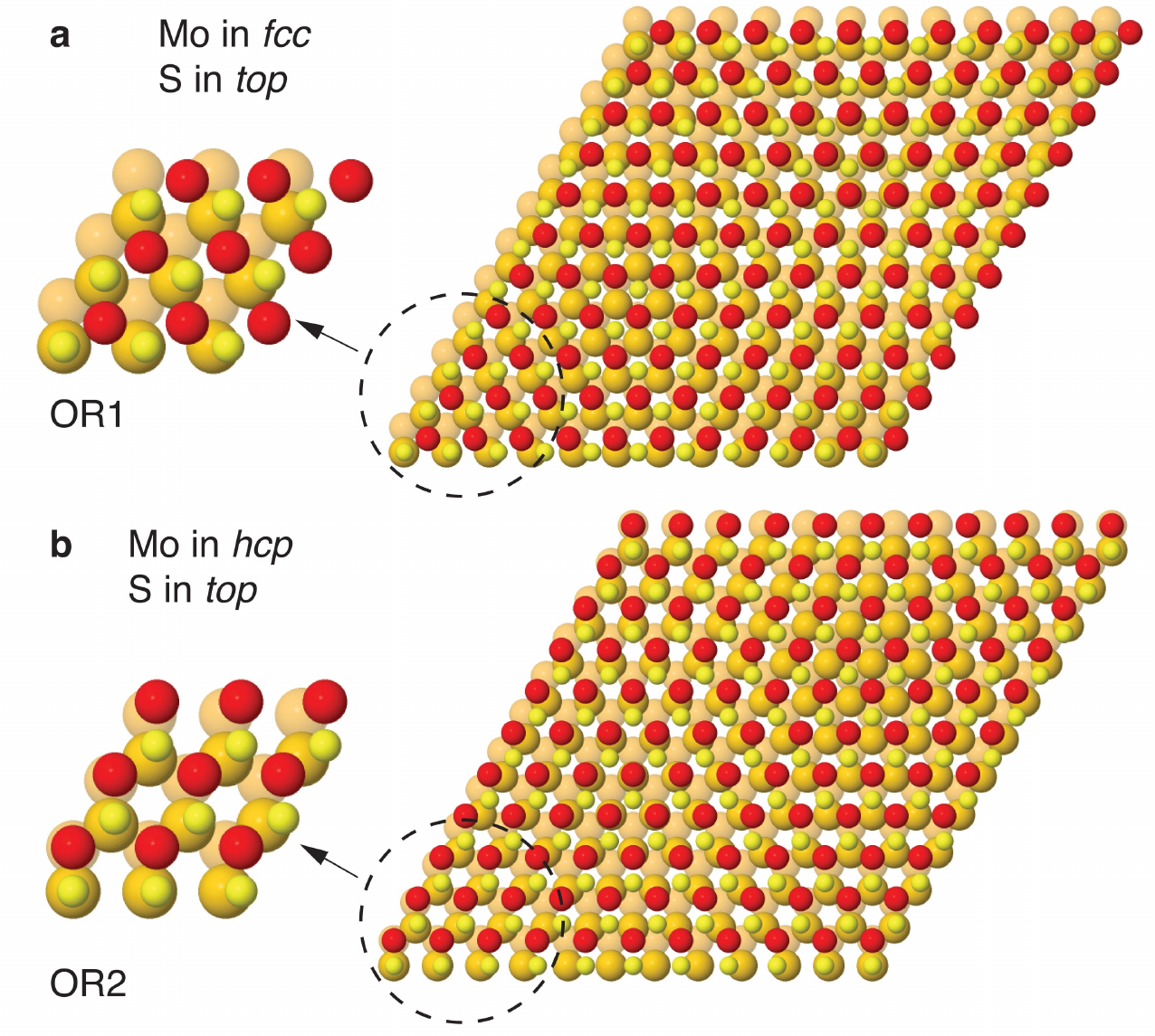}
\caption{$10\times10$-MoS$_2$/$11\times11$-Au(111) superstructure. For S adsorbed atop position, sketches on the left show the adsorption configuration of bonding regions (a) Mo in \textit{fcc} and (B) Mo in \textit{hcp} sites. Ball color code: red (Mo), yellow (S),
orange (Au surface layer) and amber (Au second layer).}
\label{registry}
\end{center}
\end{figure}
The $10\times10$ MoS$_2$ superstructure on $11\times11$ Au unit cell for the two mirror domain orientations of MoS$_2$ on Au, are shown in figure \ref{registry}. Theoretical calculations by A. Bruix et al. \cite{Bruix_2015} showed that the most favorable configuration for small 2D MoS$_2$ clusters on Au(111) is with the S atoms in atop position. Looking at the left corner of the moir\'{e} unit cell displayed in Figure \ref{registry}a and b, when the S atoms are in atop position, the Mo atoms can go in the three-fold \textit{fcc} (OR1) or \textit{hcp} (OR2) position. By combining the outcomes of the analysis of the XPD patterns stemming from Mo $3d$ and S $2p$ displayed in the main text (Figure 3) with the pattern acquired on the bulk component of Au $4f$, we can conclude that MoS$_2$ is adsorbed on Au(111) with the OR1 configuration, having thus Mo in the \textit{fcc} position. 

\subsection{R-factor analysis of the XPD patterns}
\begin{figure}[t]
\begin{center}
 	\includegraphics[width=0.8\textwidth]{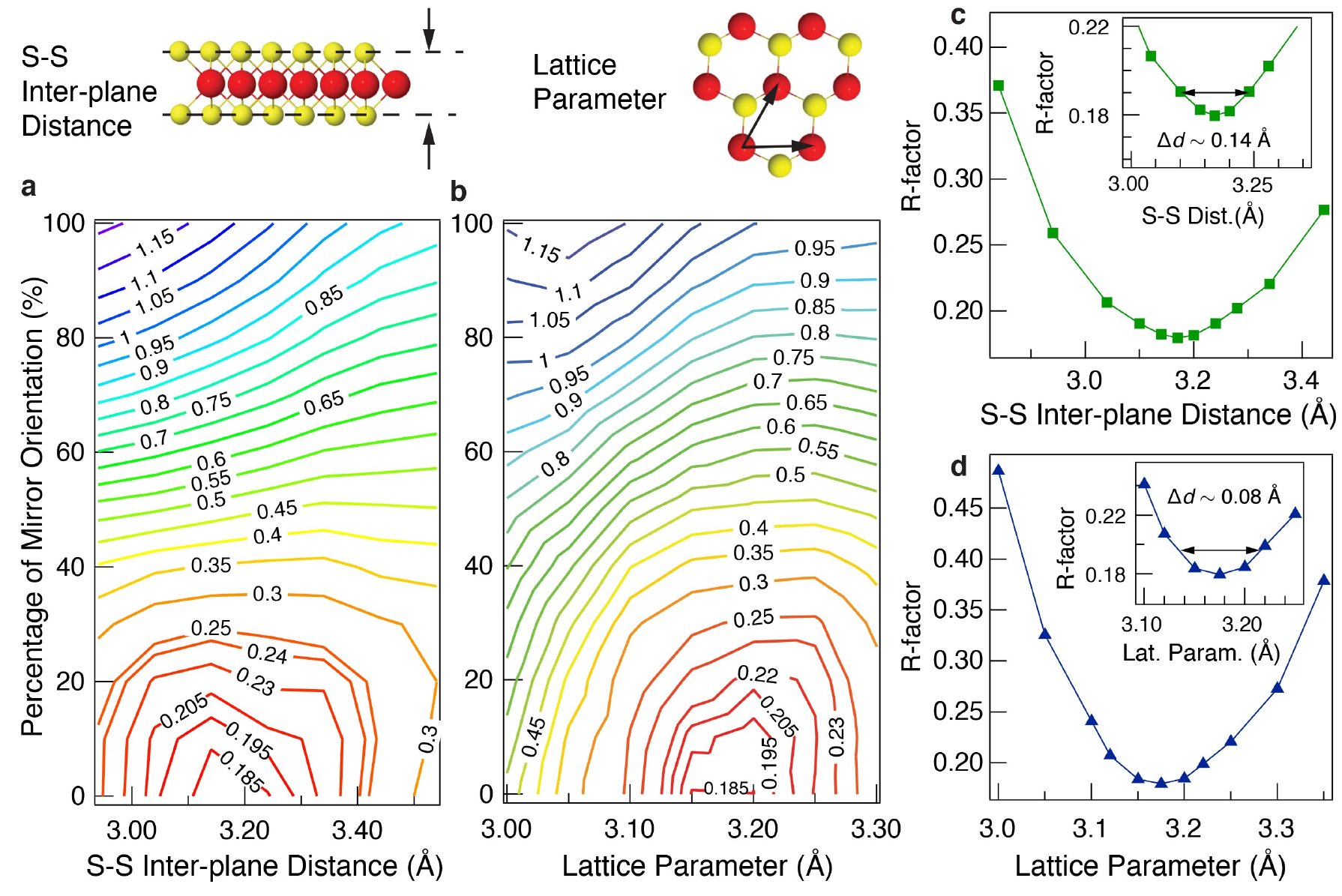}
 	\caption{Determination of MoS$_2$ structure by means of XPD. Contour plots reporting the R-factor as a function of the S-S inter-plane distance and of the percentage of mirror orientation with a lattice parameter of 3.16~\AA (a) and as a function of the lattice parameter and of the percentage of mirror orientation, with an inter-plane separation of 3.17~\AA (b). (c) and (d), R-factor plotted vs the S-S inter-plane distance and the lattice parameter, respectively, calculated for the singly-oriented MoS$_2$ layer. The insets report a detail around the minimum of the R-factor and show the error for each parameter obtained from the confidence interval, as described in the methods section.}
 	\label{contour}
	\end{center}
 \end{figure}

The structural conformation of the SL MoS$_2$ was determined by comparing experimental and simulated XPD patterns, aiming to minimise the Reliability factor (R)\cite{xpd_rfac}. Systematic multiple scattering simulations of the Mo~$3d_{5/2}$ core level were performed as a function of the MoS$_2$ lattice parameter, of the S-S inter-plane distance and of the percentage of the main (sketched in Figure \ref{contour}) and of the mirror orientation in the layer. The initial guess for the values of the lattice parameter (3.16~\AA) and for the S-S inter-plane lattice distance (3.17~\AA) were assumed from results reported for the characteristic sizes of the MoS$_2$ lattice \cite{holinski_1972}. 
In Figure \ref{contour}a and b we show contour plots reporting the R-factor as a function of the percentage of mirror orientation and of the S-S inter-plane distance and lattice parameter, respectively. A clear  minimum for the R-factor is observed when only the main orientation is included in the simulation for both plots. 

Figure \ref{contour}\textbf{c} and \textbf{d} report plots of the R-factor vs the S-S distance and the lattice parameter, respectively, for the singly-oriented layer (0\% of the mirror orientation in Figure \ref{contour}a and b). The minimum R-factor is obtained for a S-S inter-plane distance of $3.17\pm0.07$~\AA~and for a lattice parameter of $3.17\pm0.04$~\AA. The errors on these dimensions were obtained through the confidence interval $\Delta R_{min}=0.0136$, as described in the Methods section \cite{Pendry:1980aa}. These optimised distances were used to compute the diffraction patterns for S~$2p$ (Figure 1\textbf{g} and \textbf{h} in the main text), yielding R-factors of 0.25 and 0.15 for the upper and lower S atoms, respectively. 
%


\end{document}